# Large-Scale DFT Methods for Calculations of Materials with Complex Structures

Ayako Nakata[1,2]*, David R. Bowler[1,3,4], and Tsuyoshi Miyazaki[1]

[1]*International Centre for Materials Nanoarchitectonics (WPI-MANA), National Institute for Materials Science (NIMS), 1-1 Namiki, Tsukuba, Ibaraki 305-0044, Japan*
[2]*PRESTO, Japan Science and Technology Agency (JST), Kawaguchi, Saitama 333-0012, Japan*
[3]*London Centre for Nanotechnology, University College London, 17-19 Gordon St., London WC1H 0AH, United Kingdom*
[4]*Department of Physics & Astronomy, University College London, Gower St., London WC1E 6BT, United Kingdom*

Large-scale density functional theory (DFT) calculations provide a powerful tool to investigate the atomic and electronic structure of materials with complex structures. This article reviews a large-scale DFT calculation method, the multi-site support function (MSSF) method, in the CONQUEST code. MSSFs are linear combinations of the basis functions which belong to a group of atoms in a local region. The method can reduce the computational time while preserving accuracy. The accuracy of MSSFs has been assessed for bulk Si, Al, Fe and NiO and hydrated DNA, which demonstrate the applicability of the MSSFs for varied materials. The applications of MSSFs on large systems with several thousand atoms, which have complex interfaces and non-periodic structures, indicate that the MSSF method is promising for precise investigations of materials with complex structures.

## 1. Introduction

Atomic structure, electronic structure and the properties of materials correlate with each other strongly. The relationship of materials properties with local structures such as point defects in crystals and interatomic distances in glassy materials have been investigated widely for many years. Recently, wider structures ordered on the nanoscale, such as ring structures in amorphous glass,[1] ionic and molecular positions in biomaterials,[2] interfaces between metallic nanoparticle catalysts and substrates,[3] and composite dopants and defects in crystals,[4] have also been focused on as hyper-ordered structures, which could have significant influence on materials properties. The recent improvements in experimental measurements and computation techniques have enabled us to investigate such nano-scale complex structures.

For computation, density functional theory (DFT) calculations have been a powerful tool to investigate atomic and electronic structures of condensed-phase materials and molecules.[5] However, because of the high computational cost of conventional DFT calculation methods, the system size treated by DFT has been limited, up to a thousand atoms in most cases. Therefore, efficient DFT calculation methods are desirable to treat nano-scale structures. Several methods have been proposed to overcome the size limitation of DFT calculations, and we have also proposed the large-scale DFT code, CONQUEST.[6–10] There are two important methods which enable us to perform large-scale calculations with CONQUEST, the linear-scaling order-$N$, or $O(N)$, method[7,8] and the multi-site support function method.[11,12] Systems as large as one million atoms can be treated by the $O(N)$ method with a massively parallel supercomputer.[9,13] The MSSF method can be used not only to decrease the computational cost but also to improve the computational accuracy, and also enables us to perform stable large DFT calculations on metallic systems.

In the next section, we first briefly review several large-scale DFT calculation methods, and then provide more details of our large-scale calculation techniques in the CONQUEST code, especially the MSSF method. In Sect. 3, by showing several examples, we demonstrate the applicability of the MSSF method to large systems with complex structures. The final section provides the conclusion.

## 2. Large-Scale DFT Calculation Methods

### 2.1 Large-Scale DFT Calculations

The DFT total energy $E$ is a functional of electron density, which is the diagonal part of the density matrix. The density matrix can be written in terms of the Kohn-Sham (KS) one-electron

orbitals $\psi$,[14,15)]

$$\rho(\mathbf{r},\mathbf{r}') = \sum_n f_n \psi_n(\mathbf{r})\psi_n(\mathbf{r}')^*, \qquad (1)$$

where $n$ runs over KS orbitals and $f_n$ is the occupation number of $n$th KS orbital. The KS orbitals $\psi$ and their energies $\varepsilon$ are calculated as the eigenvectors and eigenvalues of the KS equation,

$$H^{KS}\psi_n(\mathbf{r}) = \left[-\frac{\hbar}{2m}\nabla^2 + V_{\text{ext}}(\mathbf{r}) + V_{\text{H}}(\mathbf{r}) + V_{\text{XC}}(\mathbf{r})\right]\psi_n(\mathbf{r}) = \varepsilon_n \psi_n(\mathbf{r}). \qquad (2)$$

In the KS Hamiltonian $H^{KS}$, the kinetic energy and the external potential $V_{\text{ext}}$ are one-electron terms, and the Hartree potential $V_{\text{H}}$ and the exchange-correlation potential $V_{\text{XC}}$ are two-electron terms. The computational cost to solve the KS equation with conventional calculation methods such as direct diagonalization scales cubically with respect to the number of atoms $N$ in a target system. This cubic scaling limits the system size for practical DFT calculations.

To overcome this limitation, several large-scale DFT calculation methods have been proposed. Since there are several review papers already which cover large-scale DFT calculation methods,[16,17)] here we briefly introduce the methods focusing on two key points, "how to solve the KS equation" and "what kind of functions are used to express the KS orbitals". The parallelization efficiency of DFT codes, which we do not describe here, is also very important.

For linear scaling solution of the KS equation for large systems, there are several methods such as the divide-and-conquer (DC) method,[18–20)] the orbital minimization method (OMM),[21,22)] the density matrix minimization (DMM) method,[23,24)] and the fragment molecular orbital (FMO) method.[25)] In the DC method, the target system is divided into small subsystems with some buffer regions whose electronic structures are calculated exactly, and then the subspace density matrices are combined to construct the density matrix of the whole target system. The DC method is used in many codes such as SIESTA[26,27)] and OpenMX.[28)] FMO is similar to the DC method, but the division is based on molecular fragments in proteins and biomolecules. The total energy of the whole system is calculated from the energy of fragments and pairs of fragments without solving for the molecular orbitals (MOs) of the whole system. Recently, a method to obtain molecular orbitals of the whole system from subspace MOs (FMO-LCMO) has been also proposed.[29)] FMO is implemented in GAMESS[30)] and ABINIT-MP.[31)] OMM and DMM are methods which minimize the total energy variationally as a functional of orbitals or density matrices, instead of solving KS equations directly. OMM is used in FEMTECK[32)] and SIESTA, and DMM is used in CONQUEST and ONETEP.[33)] There are also iterative calculation methods such as the Fermi-operator expansion method[34)] in BigDFT[35)] and the second-order trace-correcting (TC2) method[36] in ErgoSCF.[37] CP2K also uses an iterative $O(N)$ method.[38]

For the functions to represent the KS orbitals, plane-wave basis functions have been often used with the periodic boundary condition for solids and surfaces, while Gaussian- or Slater-type atomic-orbital (AO) basis functions have been popular for isolated systems such as molecules and clusters. For large-scale calculations, using local orbital functions to express the KS orbitals is crucial. Pseudo-atomic orbitals (PAOs) are atomic orbital functions constrained to go to zero at a cutoff, which avoids having long-range tails. PAOs are often used in large-scale DFT codes such as SIESTA,[39,40] OpenMX,[41,42] and CONQUEST.[43,44] There are also basis functions defined on a regular grid similarly to plane waves, which can be systematically converged, such as B-spline functions in CONQUEST,[45] periodic cardinal sine functions in ONETEP,[46] and wavelet functions in BigDFT.[47] CP2K uses Gaussian functions to describe KS orbitals and represents the electron density on a grid .[48] There are also codes using the finite difference method, for example RSDFT[49] and PARSEC.[50]

## 2.2 Large-Scale DFT Calculations in CONQUEST

In this subsection, we explain more details about our large-scale DFT code CONQUEST. In CONQUEST, the density matrix in Eq. (1) is expressed by using local orbital functions $\phi$, called "support functions", so that the density matrix **K** is represented in the support function basis,

$$\rho(\mathbf{r},\mathbf{r}') = \sum_{i\alpha,j\beta} \phi_{i\alpha}(\mathbf{r}) K_{i\alpha,j\beta} \phi_{j\beta}(\mathbf{r}')^*, \qquad (3)$$

where $\alpha$ and $\beta$ are the indices of the support function and $i$ and $j$ are the indices of the atoms which the support functions belong to. The support functions are linear combinations of given basis functions,

$$\phi_{i\alpha} = \sum_{\mu} c_{i\alpha,i\mu} \chi_{i\mu}, \qquad (4)$$

where **c** is the linear-combination coefficients and $\chi$ is the $\mu$th basis functions of atom $i$.

Two kinds of basis functions, B-spline (blip) functions[45] and PAO functions,[43,44] are available in CONQUEST. Blip functions are finite-element functions akin to plane-wave basis functions, i.e., the accuracy of the support functions can be improved systematically by making the blip functions finer. However, the optimization of the linear-combination coefficients for fine blip functions can be computationally expensive. On the other hand, the computational cost of PAOs is much cheaper than blip functions. PAOs are atomic orbital functions consisting of numerical radial functions $R$ and spherical harmonic functions $Y$,

$$\chi_{i\mu}(\mathbf{r}) = \chi_{i\zeta lm}(\mathbf{r}) = R_{i\zeta l}(r) Y_{lm}(\Omega). \qquad (5)$$

$\zeta$ is the index of the radial function, and $l$ and $m$ are the azimuthal and magnetic quantum numbers. $\Omega$ is the solid angle of $\mathbf{r}$. The accuracy of support functions with PAOs is generally improved by using multiple radial functions for each $Y$ (i.e., multiple-$\zeta$ functions) and adding $Y$ functions with larger angular-momentum numbers to represent polarization (i.e., polarization functions), although systematic improvement is not guaranteed.

When we use PAOs as the support functions without any contraction, which are called "primitive" PAOs, the coefficients $c$ in Eq. (4) are 1 or 0. When we contract multiple-$\zeta$ PAOs to smaller number of support functions, the $c$ values are optimized numerically by, for example, the conjugate gradient method. The number of support functions can be reduced to the single-$\zeta$ (SZ) size for each angular momentum functions by contraction, keeping the spatial symmetry of the primitive PAOs. For example, the primitive triple-$\zeta$ (TZ) PAOs and triple-$\zeta$ plus triple polarization (TZTP) PAOs are contracted to SZ- and SZP-sized support functions, respectively. Since the computational cost scales cubically to the number of support functions in both the exact diagonalization and the $O(N)$ calculations, it is crucial for large systems to reduce the number of support functions by the contraction. CONQUEST uses norm-conserving pseudopotentials[51–53] so that the support functions are used to describe only the valence electrons.

For the KS equation solver, CONQUEST supports both exact diagonalization and the DMM to optimize electron density. With the diagonalization method, the computational cost scales cubically with the size of the electronic Hamiltonian, i.e., the number of atoms $N$. On the other hand, the computational cost of DMM is linear with the system size. In DMM, the electronic structure is optimized by minimizing the total energy $E$ with respect to the auxiliary density matrix $\mathbf{L}$,

$$\frac{\partial E}{\partial \mathbf{L}} = 6(\mathbf{LSH} - \mathbf{HLS}) - 4(\mathbf{LSLSH} + \mathbf{LSHSL} + \mathbf{HSLSL}). \qquad (6)$$

$$\mathbf{K} = 3\mathbf{LSL} - 2\mathbf{LSLSL}, \qquad (7)$$

avoiding cubic-scaling exact diagonalization. The electronic KS Hamiltonian matrix $\mathbf{H}$ and the overlap matrix $\mathbf{S}$ are sparse because they are based on local orbital support functions. We need to set a spatial cutoff for $\mathbf{L}$, so that $\mathbf{L}$ also becomes a sparse matrix. It is guaranteed that the total DFT energy becomes equal to that by the exact diagonalization when we increase the $\mathbf{L}$ cutoff to be infinite.[54] Thus, Eqs. (6) and (7) are sparse-matrix multiplications whose computational cost are linear to the matrix size (i.e., system size).

## 2.3 Multi-Site Support Functions in Conquest

### 2.3.1 Method

Since the computational cost of both the exact diagonalization and the $O(N)$ method scales cubically to the number of the support functions, reduction of the number of support functions is crucial for the large-scale calculations. To reduce the number of support functions without losing accuracy, we have recently proposed the multi-site support functions (MSSFs).[11,12] MSSFs are defined as

$$\phi_{i\alpha} = \sum_{k}^{neighbors} \sum_{\mu \in k} c_{i\alpha,k\mu} \chi_{k\mu}, \qquad (8)$$

where $\mu$ runs over the PAOs of atom $k$ which are the neighbor atoms of atom $i$ within the cutoff distance $r_{MS}$. In Eq. (8), the MSSFs of atom $i$ consist of not only the PAOs of atom $i$ itself but also the PAOs of atom $i$'s neighbour atoms $j$, like local MOs, while the conventional support functions in Eq. (4) consist only of the PAOs of atom $i$. Since MSSFs are like local MOs, the number of MSSFs can be reduced to SZ size, which is enough to represent the ground-state accurately. For example, the SZ size for a Si atom is four, so that double-$\zeta$ plus polarization (DZP) PAOs of Si consisting of (2s, 2p, d) = 13 functions are contracted to four MSSFs, while TZTP PAOs consisting of (3s, 3p, 3d) = 27 functions are also contracted to four MSSFs. Although the numbers of the constructed MSSFs are the same, the accuracy of MSSFs from TZTP PAOs is higher than those from DZP because the number of degrees of freedom of **c** in the MSSFs is larger. This means that we can improve the accuracy of MSSFs by increasing the number of primitive PAOs without increasing the number of MSSFs. The computational cost to determine **c** increases as the number of the original PAOs increases, but the computational cost is not dominant when the whole system size is large, as shown in Sect. 3.1.1. Therefore, the MSSF method has advantages both for computational efficiency and accuracy. The **H** and **S** matrices for the whole system are reconstructed from PAO basis to MSSF basis by sparse matrix multiplication as

$$H_{i\alpha,j\beta} = \sum_{k,\mu} \sum_{p,\nu} c_{i\alpha,k\mu} H_{k\mu,p\nu} c^{*}_{p\nu,j\beta}. \qquad (9)$$

Ideally, the linear-combination coefficients **c** are optimized numerically.[12] Since the number of the coefficients of MSSFs in Eq. (8) is larger than that of the conventional onsite support functions in Eq. (4), starting the optimization from accurate initial values is desirable. The local filter diagonalization (LFD) method by Rayson et al.[55,56] is a powerful way to construct accurate initial values from localized occupied MOs in subspaces. The LFD method is based on local orbitals $\phi$ which can express the occupied KS eigenstates accurately,

$$|\phi_{i\alpha}\rangle = \sum_n |\psi_n\rangle f(\varepsilon_n)\langle\psi_n|t_{i\alpha}\rangle, \qquad (10)$$

where **t** is the $\alpha$th trial vector localized on atom $i$ and $f(\varepsilon_n)$ is the Fermi-Dirac function. In the LFD method, $\psi_n$ is replaced with subspace MOs around atom $i$ to restrict the range of $\phi_{i\alpha}$. To obtain the subspace MOs, the subspace for each atom $i$ is first defined with a cutoff range $r_{\text{LFD}}$. Then the electronic Hamiltonian and the overlap matrix for the subsystem, $\mathbf{H}_{\text{sub}}$ and $\mathbf{S}_{\text{sub}}$, are constructed to calculate the eigenvectors (i.e., subspace MOs) $\mathbf{C}_{\text{sub}}$ and eigenvalues $\varepsilon_{\text{sub}}$. Using the subspace MOs, $\phi$ is calculated as the linear-combination of the PAOs in the subspace with the linear-combination coefficients **c**,

$$\mathbf{c} = \mathbf{C}_{\text{sub}} f(\varepsilon_{\text{sub}}) \mathbf{C}_{\text{sub}}^{\text{T}} \mathbf{S}_{\text{sub}} \mathbf{t}. \qquad (11)$$

In Eq. (11), $f(\varepsilon_{\text{sub}})$ is calculated with the chemical potential of the subspace, which reduces the influence from the unphysical MOs in high-energy unoccupied region.

The calculation accuracy with $\phi$ will depend on the choice of **t**. We have chosen **t** from the primitive PAOs, but our previous study showed that the dependence is not very significant.[11] Note that $r_{\text{LFD}}$ should be equal to or larger than $r_{\text{MS}}$. Once **c** is determined, the SCF calculations of the whole systems can be performed with the matrices transformed as in Eq. (9), and $\rho$ is obtained for the **c**. Here, we can update the subsystem **H** matrices and construct new **c** by using the updated $\rho$. Thus, the update of $\rho$ and **c** is iterated until self-consistency of $\rho$ is reached. Since this update procedure is not variational, the calculated energy sometimes fluctuates, especially when $r_{\text{MS}}$ is small.[12]

By optimizing **c** numerically subsequent to the LFD calculation, not only is the accuracy of MSSFs improved but also the energy becomes variational.[12] The numerical optimization of **c** is performed with the energy gradient with respect to **c**,

$$\frac{\partial E}{\partial c_{i\alpha,k\mu}} = \frac{\partial E}{\partial \phi_{i\alpha}} \frac{\partial \phi_{i\alpha}}{\partial c_{i\alpha,k\mu}} = \frac{\partial E}{\partial \phi_{i\alpha}} \chi_{k\mu}. \qquad (12)$$

The gradient with respect to $\phi_{i\alpha}$ is calculated as

$$\frac{\partial E}{\partial \phi_{i\alpha}} = 4\sum_\beta \left[ K_{\alpha\beta}\hat{H} + G_{\alpha\beta} \right] \phi_{j\beta}, \qquad (13)$$

where **G** is given as the energy-weighted density matrix

$$G_{\alpha\beta} = \sum_n f_n \varepsilon_n u_{n\alpha} u_{n\beta}^* \qquad (14)$$

in diagonalization calculations (**u** is the KS coefficients in the support function basis) and as

$$G_{\alpha\beta} = 3(LHL)_{\alpha\beta} - 2(LSLHL + LHLSL)_{\alpha\beta} \qquad (15)$$

in $O(N)$ calculations. Detailed derivations of Eqs. (13) – (15) are provided in our previous papers.[7,57]

### 2.3.2 Investigation of Accuracy and Efficiency

In this subsection, the accuracy of MSSF is investigated by checking the $r_{MS}$ dependence and the effect of the numerical optimization. All calculations shown in the present paper were performed with the exact diagonalization, not with the $O(N)$ method to avoid the error coming from the cutoff of **L**, and $r_{LFD}$ was set to be equal to $r_{MS}$. More detailed information about the computational conditions such as the number of k-points are found in the references.

First, we investigate the accuracy of the calculated energies for bulk Si and Al systems with the local density approximation (LDA)[58] exchange-correlation functional. Figure 1 shows the deviations of the DFT total energies by the MSSFs from those by the primitive PAOs.[11] The TZP PAOs (3s, 3p, d) consisting of 17 functions are contracted to four MSSFs for both Si and Al atoms. The energy deviations of the MSSFs decrease exponentially as $r_{MS}$ increase. It is worth noting that the energy of the bulk Al converges smoothly with respect to the spatial cutoff $r_{MS}$ although the bulk Al is metallic; this is because the cutoff $r_{MS}$ is introduced only to the support functions, not to the wave function of the whole system.

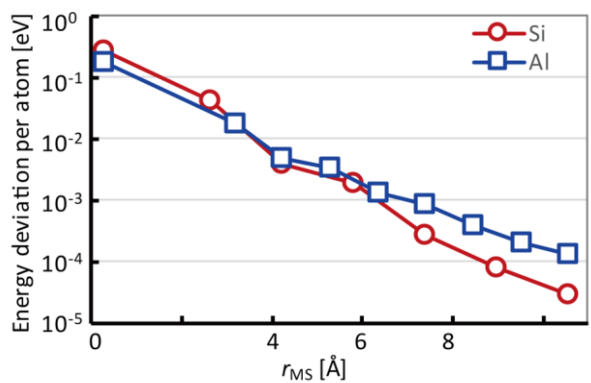

**Fig. 1**. Deviation of total DFT energy per atom of bulk Si and Al of multi-site support functions with respect to the multisite range $r_{MS}$ from the DFT energy of primitive TZP PAOs. (Data taken with permission from Ref. 11. © 2014 American Chemical Society.)

The comparison of the energy-volume (E-V) curves of bulk Si with several $r_{MS}$ is shown in Fig. 2.[12] The MSSFs are constructed from triple-$\zeta$ plus double polarization (TZDP) PAOs. The results of the MSSFs constructed only by LFD and those of the MSSFs by LFD and subsequent numerical optimization are compared in the figure. The lattice constants $a_0$ calculated by fitting the E-V curve with the Birch-Murnanghan equation are summarized in Table I. When only LFD is used to determine **c**, MSSFs with $r_{MS}$ = 2.6 Å, which contain only the nearest neighbor atoms in $r_{MS}$, show an error of about 1.0 %, while MSSFs with larger $r_{MS}$, 4.2 Å and 8.5 Å, show much smaller errors, 0.2 % and 0.0 %, respectively. $r_{MS}$ = 4.2 Å contains up to the second-nearest neighbor atoms. The numerical optimization improves the accuracy of the MSSFs: when we optimize **c** after the LFD method, even the error of MSSFs with $r_{MS}$ = 2.6 Å is small, only 0.2 %.

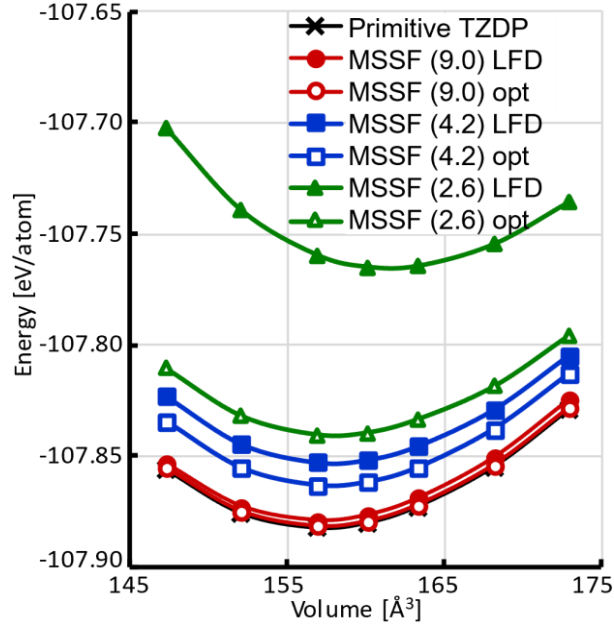

**Fig. 2**. Energy-volume curve of bulk Si calculated using multi-site support functions (MSSFs) with LFD, MSSFs with numerical optimization, and primitive TZDP PAOs. The multi-site ranges $r_{MS}$ [Å] are shown in parentheses. (Reproduced from Ref. 12 with permission from the PCCP Owner Societies.)

**Table I**. Lattice constants $a_0$ of crystalline Si calculated with multi-site support functions (MSSFs) using the local filter diagonalization method (LFD), MSSFs using LFD plus numerical optimization, and primitive TZDP PAOs. The percentage deviations from the result by TZDP are also shown. (Reproduced from Ref. 12 with permission from the PCCP Owner Societies.)

|  | $a_0$ [Å] | | %Δ from primitive TZDP | |
| --- | --- | --- | --- | --- |
|  | LFD | LFD + opt | LFD | LFD + opt |
| MSSF ($r_{MS}$ = 2.6 Å (5.0 bohr)) | 5.447 | 5.406 | 1.0 | 0.2 |
| MSSF ($r_{MS}$ = 4.2 Å (8.0 bohr)) | 5.403 | 5.400 | 0.2 | 0.1 |
| MSSF ($r_{MS}$ = 9.0 Å (17.0 bohr)) | 5.393 | 5.395 | 0.0 | 0.0 |
| primitive TZDP |  | 5.395 |  |  |

To check the accuracy of atomic forces, the energies and forces have been investigated for a distorted benzene molecule in which a C-H pair has been shifted away from the center of the benzene ring by 0.5 Å. Table II shows the differences of the energies and the maximum forces by the MSSFs with several $r_{MS}$ from those by primitive DZP PAOs. It was clearly found that the increase of $r_{MS}$ improves the accuracy of both energies and forces. When optimizing MSSF coefficients, even MSSFs with $r_{MS}$ = 1.6 Å provide comparable accuracy with DZP PAOs for forces, within 0.05 eV/Å.

**Table II**. Differences of the total energies and forces of a distorted benzene molecule calculated using multi-site support functions (MSSFs) with the local filter diagonalization method (LFD) and MSSFs with LFD plus numerical optimization, from those by the primitive TZDP PAOs. (Part of this table is reproduced from Ref. 12 with permission from the PCCP Owner Societies.)

|  | Energy difference [eV] | | Force difference [eV/Å] | |
| --- | --- | --- | --- | --- |
|  | LFD | LFD + opt | LFD | LFD + opt |
| MSSF ($r_{MS}$ = 1.6 Å (3.0 bohr)) | 7.312 | 0.110 | -12.499 | 0.038 |
| MSSF ($r_{MS}$ = 2.6 Å (5.0 bohr)) | 0.108 | 0.013 | 0.017 | -0.005 |
| MSSF ($r_{MS}$ = 4.2 Å (8.0 bohr)) | 0.002 | 0.001 | -0.015 | 0.004 |
| primitive DZP | -1019.943 | | -0.642 | |

Next, the accuracy of the electronic structure was investigated by checking the density of states (DOS) of a hydrated DNA system with 3,088 atoms.[12] The PBE generalized gradient approximation (GGA) functional was used.[59] The difference of the DOS calculated with the primitive PAOs and the MSSFs are presented in Fig. 3. The difference is smaller in the occupied states and low-energy unoccupied states than in the high-energy unoccupied states. The difference in the high-energy unoccupied states is larger with smaller $r_{MS}$. The accuracy in the occupied states is improved by the numerical optimization of **c**, but the description of the unoccupied states becomes less accurate. This is because the linear-combination coefficients **c** are optimized only for the occupied states. The unoccupied states can be improved by cooperating with the Sakurai-Sugiura method,[60] which will be shown in Sect. 3.4.

The computational efficiency was also investigated for the hydrated DNA system.[12] Table III summarizes the times required for matrix construction, diagonalization and gradient calculation with respect to the coefficients **c** (i.e., Eq. (12)) with primitive DZP and MSSFs. For the matrix construction, the MSSFs require additional time to construct **c** and to reconstruct the overlap and Hamiltonian matrices in the MSSF basis. This additional time increases as $r_{MS}$ becomes larger. On the other hand, for the diagonalization, MSSFs reduces the time dramatically with any $r_{MS}$: about 600 seconds with MSSFs compared to about 12,000 seconds with primitive DZP. $r_{MS}$ does not affect the time for diagonalization because the number of MSSFs does not depend on $r_{MS}$. The times for the gradient calculations are shorter than those for the matrix constructions and the diagonalizations. Because the reduction of the diagonalization time is much larger than the increase of the times for the matrix construction and the gradient calculations, the total computational time can be reduced significantly although we need to iterate the SCF and the gradient calculations until the coefficient optimization converges. Since the number of the iterations depends on the accuracy of the initial values of the coefficients, we expect that the present optimization method will be more efficient if we can use coefficients from previous steps, such as in molecular dynamics simulations and geometry optimizations.

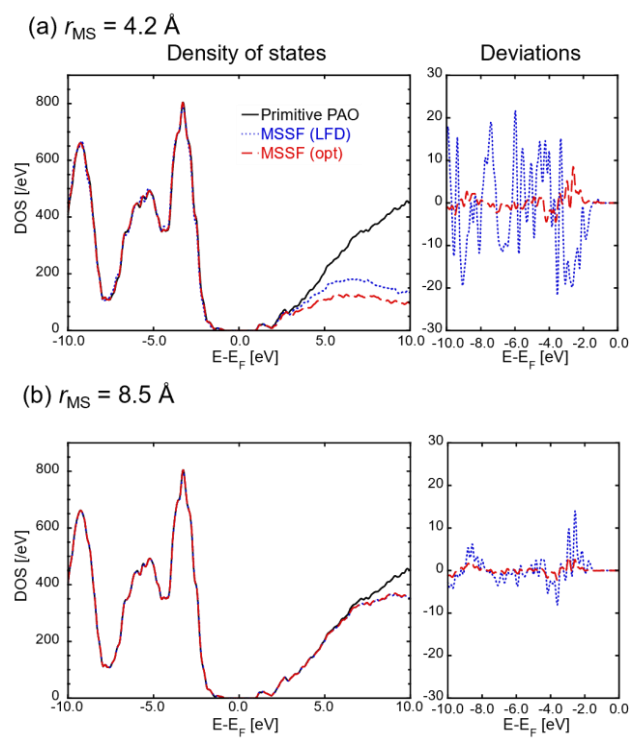

**Fig.3**. Density-of-states of the hydrated DNA system using primitive (black, solid) and multi-site support functions with local filter diagonalization (blue, dotted) and numerical optimization (red, dashed). Deviations from the primitive functions are shown for the occupied states. (Reproduced from Ref. 12 with permission from the PCCP Owner Societies.)

**Table III**. Computational times [seconds] for matrix construction, diagonalization and gradient calculation with respect to the coefficients for the hydrated DNA system by multi-site support functions (MSSFs) and primitive DZP PAOs. (Data taken from Ref. 12 with permission from the PCCP Owner Societies.)

|  | Matrix construction | Diagonalization | Gradient calculation |
|---|---|---|---|
| MSSF ($r_{MS}$ = 4.2 Å (8.0 bohr)) | 34.5 | 627.7 | 4.3 |
| MSSF ($r_{MS}$ = 5.8 Å (11.0 bohr)) | 54.4 | 583.4 | 8.6 |
| MSSF ($r_{MS}$ = 8.5 Å (16.0 bohr)) | 484.8 | 566.3 | 25.7 |
| primitive DZP | 28.0 | 12317.4 |  |

*2.3.3 Spin-Dependent MSSFs*

For the spin-polarized systems, there are two possibilities for how to determine MSSF linear-combination coefficients: determining the coefficients for spin-up and spin-down electrons individually; and using the same coefficients for spin-up and spin-down electrons by taking their average. The first method leads to spin-dependent MSSFs. Spin-dependent MSSFs are obtained by using spin-up and spin-down **C**$_{sub}$ in Eq. (11) (and subsequent numerical optimization to minimize total DFT energies with respect to both spin-up and spin-down coefficients). When we use spin-dependent MSSFs, not only the two-electron terms but also the one-electron terms in the KS Hamiltonian and the overlap matrix become spin dependent, so that additional memory for those matrices is required. The second method leads to spin-independent MSSFs so does not require the additional memory, but the accuracy might be lower than spin-dependent MSSFs.

Therefore, we compare the accuracy of spin-dependent and spin-independent MSSFs for spin-polarized systems: bulk bcc ferromagnetic Fe and cubic antiferromagnetic NiO. $r_{MS}$ is set to be 8.5 Å for both Fe and NiO. Figure 4 shows the E-V curves of the Fe and NiO, and the $a_0$ values fitted from the E-V curves are summarized in Table IV. In Fig. 4, the energy values and the curvature of the spin-dependent MSSFs are closer to those of the PAOs than spin-independent MSSFs are. $a_0$ of the spin-dependent MSSFs is closer to that of PAOs than spin-independent MSSFs for Fe, while they are comparable for NiO. The magnetic moments $\mu_B$ of bcc Fe are calculated to be 2.40, 2.26 and 2.34 by the primitive PAOs, spin-independent MSSFs and spin-dependent MSSFs, respectively. Figure 5 shows the calculated DOS of antiferromagnetic NiO. The differences of the DOS from that of PAO is also shown, in which spin-dependent MSSFs clearly reduce the differences for both spin-up and spin-down states. These comparisons indicate that the accuracy of the calculation is improved by considering spin-dependence of MSSFs.

**Table IV**. Lattice constants $a_0$ [Å] of bcc ferromagnetic Fe and cubic antiferromagnetic NiO calculated using primitive DZP PAOs, spin-independent multi-site support functions (SI-MSSF) and spin-dependent multi-site support functions (SD-MSSF).

|     | PAO   | SI-MSSF | SD-MSSF |
|-----|-------|---------|---------|
| Fe  | 2.890 | 2.876   | 2.885   |
| NiO | 8.446 | 8.449   | 8.449   |

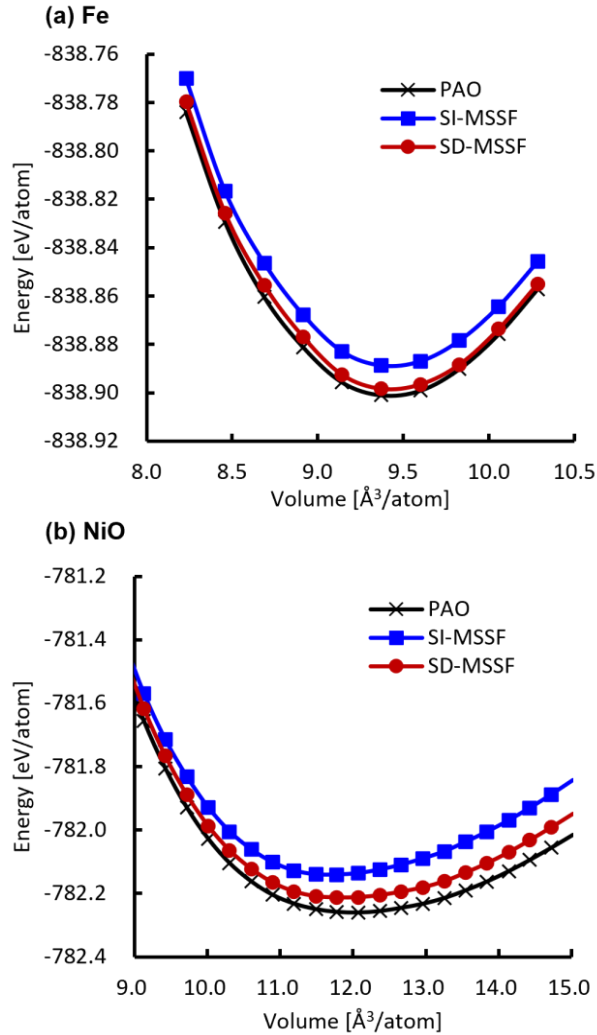

**Fig.4**. Energy-volume curve of (a) bcc ferromagnetic Fe and (b) cubic antiferromagnetic NiO calculated with primitive DZP PAOs, spin-independent multi-site support functions (SI-MSSF), and spin-dependent multi-site support functions (SD-MSSF).

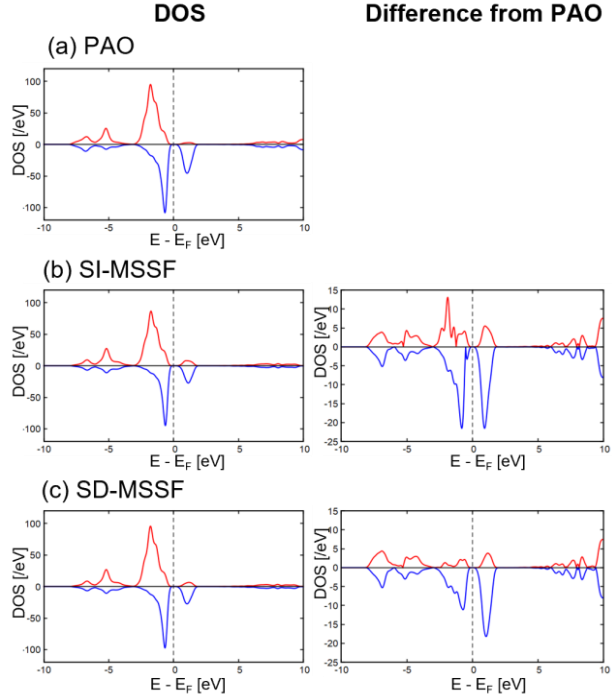

**Fig.5**. Density of states (DOS) of cubic antiferromagnetic NiO calculated with (a) primitive DZP PAOs, (b) spin-independent multi-site support functions (SI-MSSF), and (c) spin-dependent multi-site support functions (SD-MSSF). The absolute differences of the DOS from (a) are also shown for (b) and (c). Spin-up (red) and spin-down (blue) states are shown as positive and negative values, respectively.

## 3. Applications of MSSFs on Large Systems

To demonstrate the actual applicability of MSSFs to large systems, in this section we show several examples of the applications of MSSFs to large complex structures: moiré graphene on the Rh(111) surface[61]; the interfaces in YGaO$_3$[62]; and PbTiO$_3$ films on SrTiO$_3$.[63] The applications to nonperiodic systems, hydrated DNA[60] and metallic gold nanoparticles, are also shown. The LDA exchange-correlation functional is used for YGaO$_3$ and PbTiO$_3$ on SrTiO$_3$, and the PBE functional is used for graphene on Rh(111), hydrated DNA and Au nanoparticles. Norm-conserving pseudopotentials have been used in the calculations with CONQUEST. The other detailed information about the computational conditions such as the number of k-points and the spatial range of PAOs are found in the corresponding reference papers. In several examples, plane-wave calculations for comparison have been performed with the PAW pseudopotential[64] using the VASP software.[65,66]

*3.1 Graphene on Rh Surface*

There have been many reports showing the exotic properties of 2D materials. The structure of 2D materials and their electronic structures are often affected by the interactions with the substrates or interlayer interactions. The target system in Ref. 61 also shows the interesting property that a highly corrugated graphene layer grown on Rh(111) can be flattened by the intercalation of oxygen atoms. For this system, plane-wave DFT calculations demonstrated that strong interactions between the graphene layer and the substrate are decoupled when oxygen atoms are intercalated in the lowest moiré sites. Such DFT simulations have been performed to clarify the structural and electronic properties of 2D materials, but they were limited in the size of simulation cells and it was difficult to study the case of low concentration of oxygen atoms, large moiré structures, the effect of the edges and so on. It is expected that these obstacles can be overcome if accurate MSSF method can be applied to such 2D materials.

In Ref. 61, the accuracy and computational-time efficiency of the MSSF method with a system, graphene on Rh(111) surface (G/Rh), were investigated. First, the accuracies of the PAOs in the optimized structures of G/Rh and G/O/Rh having small supercells were confirmed by comparing to plane-wave basis functions. TZDP PAOs of carbon atoms, TZTP PAOs of oxygen atoms and DZP PAOs of rhodium atoms were contracted to form MSSFs with $r_{MS}$ = 8.5 Å. The DFT-D2 dispersion correction method[67] was used to consider van der Waals interaction. At first, the structural parameters, $a_0$ and bulk modulus $B_0$, of graphene and bulk Rh were compared, as shown in Table V. Figures 6(a) and 6(b) show the DOS of the G/Rh(111) with and without inserted oxygen atoms. As shown in the table and the figure, the results by the PAOs and the MSSFs can be very close to those by the plane-wave basis functions.

**Table V**. Lattice parameters $a_0$ and bulk moduli $B_0$ of graphene and bulk rhodium calculated with plane-wave basis functions and multi-site support functions. (Data taken from Ref. 61.)

|      | Graphene | Bulk rhodium | |
| --- | --- | --- | --- |
|      | $a_0$ [Å] | $a_0$ [Å] | $B_0$ [GPa] |
| PW[a] | 2.4678 | 3.7729 | 270 |
| PAO  | 2.4762 | 3.7903 | 255 |
| MSSF | 2.4767 | 3.7844 | 265 |
| Exp. | 2.46 | 3.8 | 269 |

[a] By VASP with energy cutoff 400 eV.

For large-scale calculations, the computational times by the primitive PAOs and the MSSFs for the G/Rh(111) consisting of 3,088 atoms (shown in Fig. 6(c)) are compared in Table VI. These large systems were too computationally expensive to be treated with plane-waves. The calculations were performed with the supercomputer SGI ICE X (Intel Xeon E5-2680V3 (12 cores, 2.5 GHz) × 2 and 128 GB memory per node) at NIMS. In the case of 108 processes, the computational time with the MSSFs is about 18 times shorter than that with the PAOs. Although additional time to construct the contraction coefficients **c** is needed (the time is included in "Matrix construction" in Table VI), the time for the diagonalization of the electronic Hamiltonian of the whole system, which is conventionally dominant in the total computational time, is dramatically reduced. This is because the computational cost of the diagonalization scales cubically with the number of support functions, which is reduced dramatically by the MSSF method. When the number of cores increases eight times, the computational time with MSSFs is reduced from 2156 sec. to 571 sec., almost a quarter. Although this is not ideal scaling, it demonstrates that a large speedup can still be achieved by parallelization. It should be also emphasized that large-scale DFT calculations are available with the MSSF method even for metallic systems.

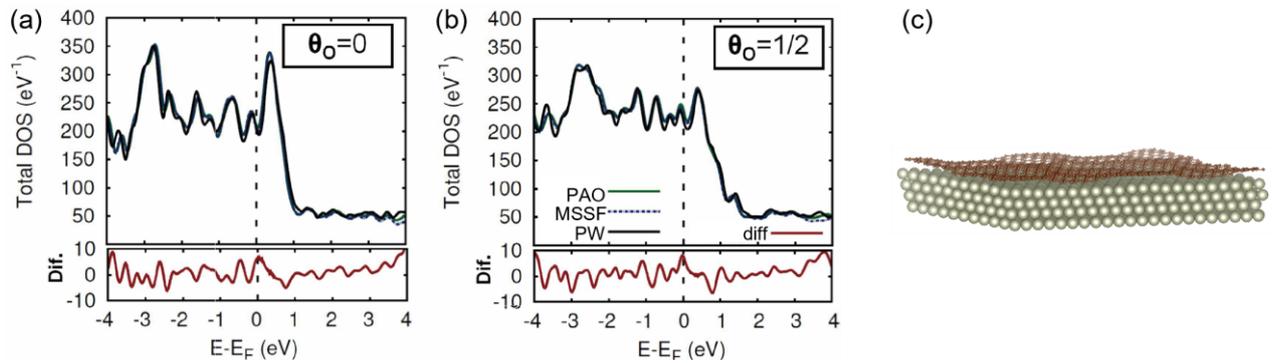

**Fig. 6**. Density of states for graphene on a Rh(111) substrate (460 atoms) with no oxygen atom ($\theta_O = 0$) and with 12 oxygen atoms in the interface ($\theta_O = 1/2$) calculated with plane waves (black), PAOs (green) and multi-site functions (blue dashed). The red lines in the lower panels represent the DOS difference between the PAOs and the MSSF calculations. Fermi-levels are set to be zero (black dashed). (c) Structure of corrugated graphene on Rh (111) with 3088 atoms. (Reproduced with permission from Ref. 61. © 2018 IOP Publishing.)

**Table VI**. Computational times of an SCF step with PAOs and MSSFs for graphene on Rh(111) surface with 3088 atoms. (Data taken from Ref. 61.)

| Function | | PAO | MSSF | MSSF |
|---|---|---|---|---|
| No. of support functions | | 54384 | 16244 | 16244 |
| No. of MPI processes | | 108 | 108 | 864 |
| No. of nodes | | 72 | 72 | 36 |
| Time [sec] | Matrix construction | 155.7 | 1455.4 | 405.9 |
| | Diagonalization | 37647.7 | 700.8 | 165.9 |
| | sum | 37803.5 | 2156.3 | 571.8 |

### *3.2 Interfaces in Ferroelectric YGaO$_3$*

The next example is an investigation of topological defects in ferroelectric YGaO$_3$.[62] Ferroelectric domain walls are attracting broad attention for next-generation nanoelectronics. Although the basic properties of simple ferroelectric domain walls can be well described by small DFT calculations, complex domain patterns could not be treated since very large supercells are needed to model the structure. The target in Ref. 62 is a vortex core at which six kinds of structural domains meet. To model the complex structure with a periodic boundary condition, two pairs of vortex/antivortex cores need to be included in the calculation cell, which contains at least about 3,600 atoms. Using the MSSF method with CONQUEST, the atomic-scale structure of the vortices and their electronic structures have been investigated. TZDP PAOs were contracted to MSSFs with $r_{MS}$ = 6.4 Å in the calculations.

Before studying the complex topologically protected vortex cores, calculations of two domain walls in the 1x12x1 supercell containing 360 atoms (shown in Fig. 7) have been performed. The accuracy of the calculations using primitive TZDP PAOs and MSSFs was confirmed by comparing to plane-wave calculations. The phase $\Phi$ and the tilt angle $Q$ (Fig. 7(a)) in the optimized structure are shown in Figs. 7(b) and 7(c). The profiles of $\Phi$ and $Q$ found by TZDP PAOs and MSSFs are almost the same and are only slightly different from those found by the plane wave calculations. The average of the formation energy of the two domain walls in the two-domain model were calculated as 15.22, 14.43 and 13.72 mJ/m$^2$ by TZDP PAOs, MSSFs and the plane waves, respectively, which are within ~3 mJ/m$^2$ (~0.1 meV/Å$^2$) difference. All these results support the robustness and accuracy of the MSSF method in this system.

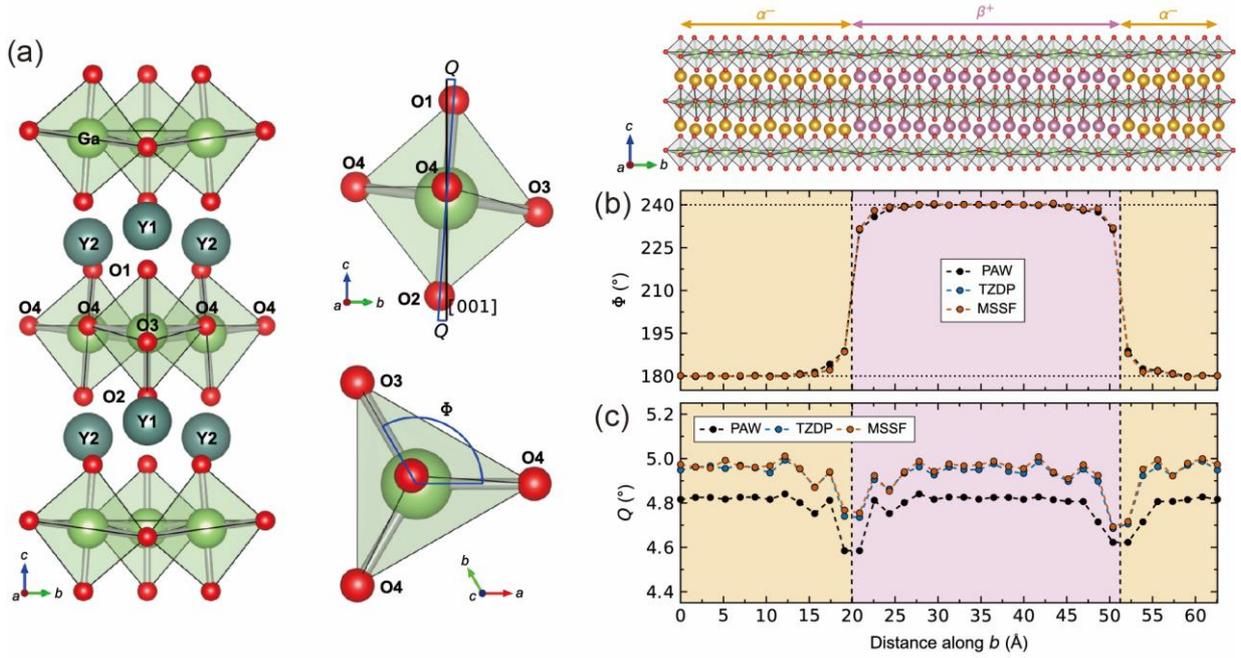

**Fig. 7.** (a) Crystal structure of ferroelectric YGaO$_3$ (*P6$_3$cm* space group). The amplitude $Q$ and phase $\Phi$ are quantified by the GaO$_5$ tilt angle relative to the [001] direction and the GaO$_5$ tilt direction projected onto the *ab* plane, respectively. (b) $\Phi$ and (c) $Q$ in the optimized structure of 360 atoms 1×12×1 supercell with two-domain patterns. (Reproduced with permission from Ref. 62. © 2020 American Physical Society.)

Then the topologically protected vortex using a 3,600-atom, 10×12×1 supercell, have been investigated with MSSFs. The model consists of two vortex/antivortex pairs, where the domain walls and the vortices are initialized in *P6$_3$/mmc* symmetry, as taken by the paraelectric phase. $\Phi$ and $Q$ in the optimized structure are visualized in Figs. 8(a) and 8(b). The optimized structure indicates that the structure around the core adopts *P3c1* symmetry. The DOS and the electronic density in the energy region around the conduction band minimum are illustrated in Figs. 8(c) and 8(d). In Fig. 8(d), the vortex core has larger amplitude than the other areas. According to the plane-wave calculations, the band gap of the YGaO$_3$ unit cell is decreased from 3.19 eV to 2.76 eV by shifting the CBM due to the symmetry change from *P6$_3$/mmc* to *P3c1*. These results suggest that the band gap of YGaO$_3$ will be reduced by the symmetry change at the vortex core. Thus, it has been demonstrated that the large-scale DFT calculations with MSSFs help us to investigate details of complex vortex structures, leading to specific electronic structures.

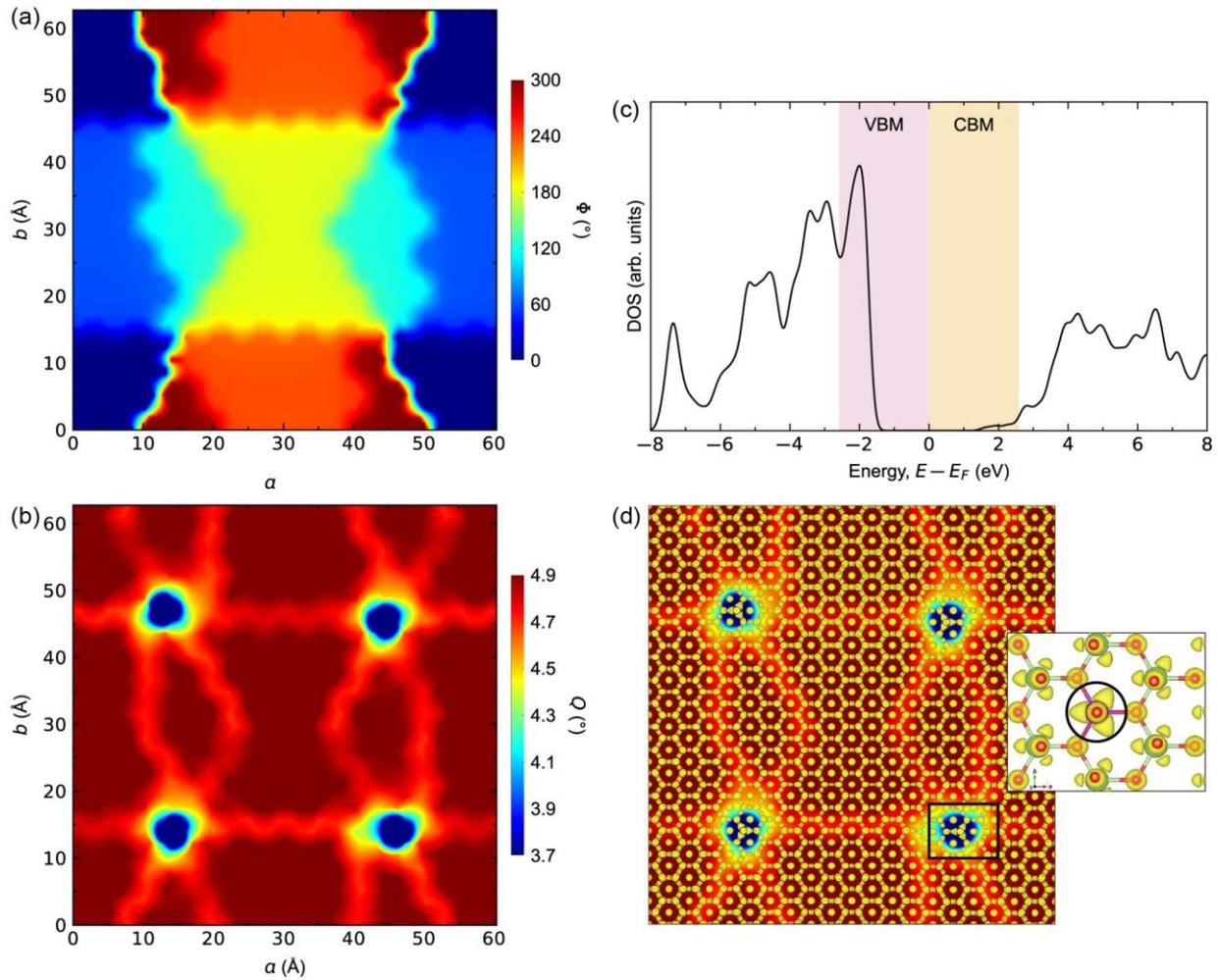

**Fig. 8**. Surface map of (a) $\Phi$ and (b) $Q$ in the optimized structure and (c) density of states and (d) electron density around the conduction band minimum (yellow region in (c)) of 3600 atoms 10×12×1 supercell with two vortex/antivortex pairs. (Reproduced with permission from Ref. 62. © 2020 American Physical Society.)

### 3.3 PbTiO₃ Films on SrTiO₃ Substrates

We have investigated a perovskite material, PbTiO$_3$ films on SrTiO$_3$ substrate.[63] Advanced deposition techniques allow the creation of thin film perovskite oxides and layered heterostructures, which demonstrate a wide variety of electrical polarization textures with possible applications in low dimensional functional devices. These textures arise from the interaction of different order parameters, notably anti-ferro-distortive (AFD) and ferroelectric (FE) distortions. At the surface of PbTiO$_3$ (PTO), antiphase rotations of the TiO$_6$ octahedra give rise to an AFD c(2×2) reconstruction. We used CONQUEST with MSSF to model films of PTO on SrTiO$_3$ (STO) with a variety of polar morphologies: paraelectric; monodomain FE in-plane

(both along (100) and (110) directions); and polydomain FE films with polarization along (001). This last morphology is particularly challenging, as it requires the simulation cell to include the substrate, the in-plane width of the domain (which increases with thickness) and a doubling along the (010) direction to allow for the AFD rotations. Most simulations of thin films use a superlattice, removing the possibility to examine the important effect of the surface on the polarization textures.

We found that seven layers of STO substrate were required to avoid any influence on the PTO; we then built cells with between one and nine layers of PTO for the polar morphologies (described above/shown in Fig. 9). We used MSSFs with $r_{MS}$ = 6.4 Å (which converged all relevant parameters) and worked with the LDA, which gives excellent values of bulk polarization for PTO. In the polydomain film, domain walls lie on PbO planes.

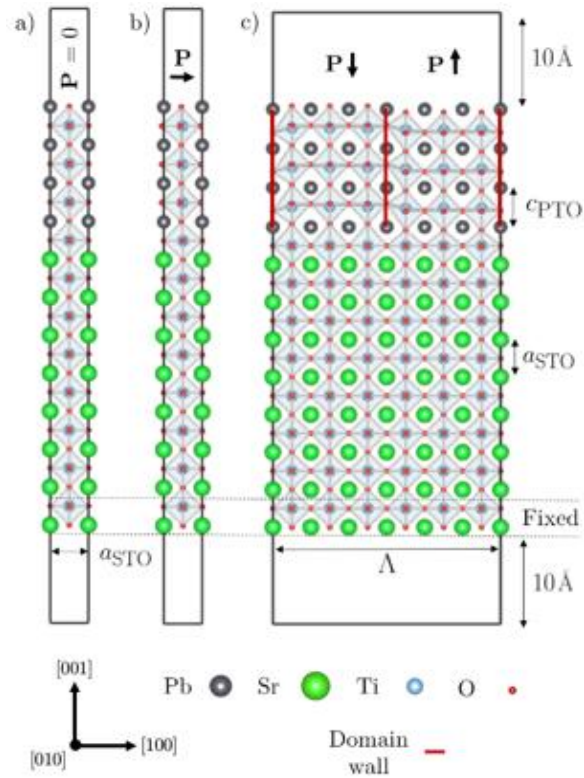

**Fig. 9**. The initial supercell configurations for the $N_z$ = 3 films before structural relaxation. Shown here are the supercells not including AFD modes. Each configuration is, however, also treated with AFD modes following the explanation in Section 3.3. a) The paraelectric supercell constrained such that spontaneous polarization cannot emerge. b) The monodomain in-plane ferroelectric case (**P** ∥ [100] is shown here, but we also treat **P** ∥ [110]) constrained such that spontaneous polarization cannot develop in the out-of-plane direction. c) The polydomain ferroelectric case with equally sized up and down domains for the ferroelectric polarization. Shown here is the Λ = 6 case. (Used with permission from Ref. 63. © 2020 The Authors.)

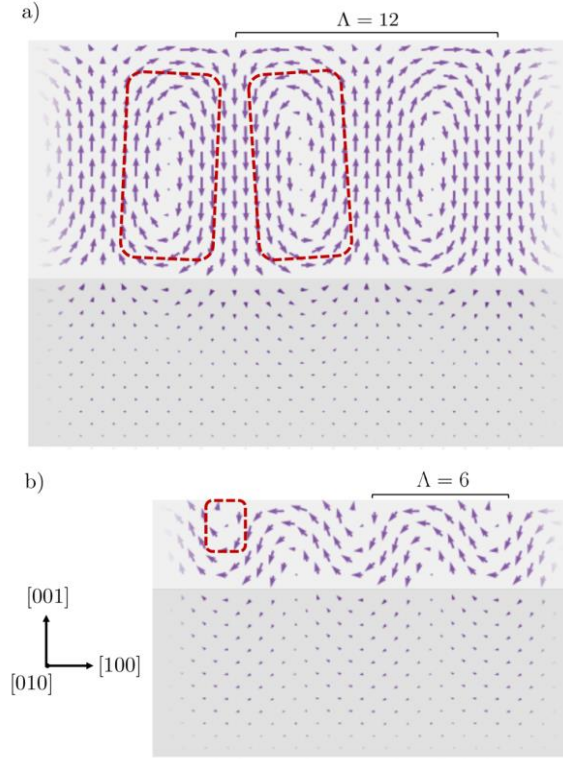

**Fig. 10**. The local polarization vector fields in the x–z plane for two film thicknesses not including AFD modes. (a) The flux-closure domains of the $N_z = 9$, $\Lambda = 12$ film. The red dashed area highlights a vortex/antivortex pair. (b) The polar wave morphology in the $N_z = 3$, $\Lambda = 6$ film. The red area indicates a cylindrical chiral bubble. (Used with permission from Ref. 63. © 2020 The Authors.)

We find that the polydomain arrangement is unstable for thicknesses of N<3, and is less stable than the monodomain (110) film with AFD distortions, though is more stable than the (100) monodomain film, up to N=5; after that point, polydomains are most stable. This fits well with experimental observations that polydomains start to be observed for N>=3. The local polarization fields for two film thicknesses are shown in Fig. 10. Thick films, such as that shown in Fig. 10(a), form flux-closure domains with clear domain walls visible. For thin films, we find a polar wave morphology with small chiral cylindrical bubbles forming at the surface. As a consequence of these detailed reconstructions, we find that the surface periodicity is not c(2×2) as would be expected from the AFD distortions alone, but instead p(2×Λ), where Λ depends on the film thickness.

The full exploration of these exotic polarization textures at thin film surfaces is only possible through the use of large-scale DFT, as enabled by MSSF in CONQUEST.

*3.4 Hydrated DNA together with Sakurai-Sugiura Method*

Large-scale calculation methods are also important to investigate non-periodic materials such as glassy materials, polymers and biomaterials. For example, in this section, MSSF have been used to model a hydrated DNA system[60], shown in Fig. 11. The solvent water molecules have been treated explicitly in the calculations, therefore the system consists of 634 atoms in the DNA, 932 hydrating water molecules and 9 Mg counterions, in total 3,439 atoms. DZP PAOs with 27,883 primitive functions have been contracted to 7,447 MSSFs. Figure 11(a) compares the DOS of the hydrated DNA calculated by the primitive PAOs and the MSSFs. As discussed in Sect. 2.3.2, the MSSFs have reproduced the DOS of the primitive PAOs with high accuracy for the occupied states but not for unoccupied states. To improve the accuracy of the unoccupied states, we have introduced the Sakurai-Sugiura method (SSM).[60,68,69] SSM is an interior eigenproblem solver for large sparse matrices, providing the eigenvalues and eigenvectors in given energy regions with high parallel efficiency.[68,69] We have performed a SCF calculation to optimize the electronic density using MSSFs and re-constructed the electronic Hamiltonian in primitive PAO basis with the optimized electronic density. Subsequently, one-shot SSM calculations have been performed for the energy region of interest. For the hydrated DNA system, we have performed the SSM calculations for the energy range [-1:1] eV with an interval of 0.027 eV (0.001 hartree). Thus, the unoccupied states of the hydrated DNA have been improved as in Fig. 11(a). Figure 11(b) shows the calculated HOMO and LUMO of the DNA system. The investigation of the KS states is important to investigate materials properties such as intramolecular electron transfers. The combination of SSM with $O(N)$ calculations in CONQUEST is also a powerful tool to investigate the electronic structure of extremely large systems. Although the $O(N)$ method itself does not provide information of KS eigenstates, SSM can provide the eigenstates (= KS states) for the $O(N)$ Hamiltonian. The KS-state calculations around the Fermi level for a system with 194,573 atoms has been achieved with this combination.[60]

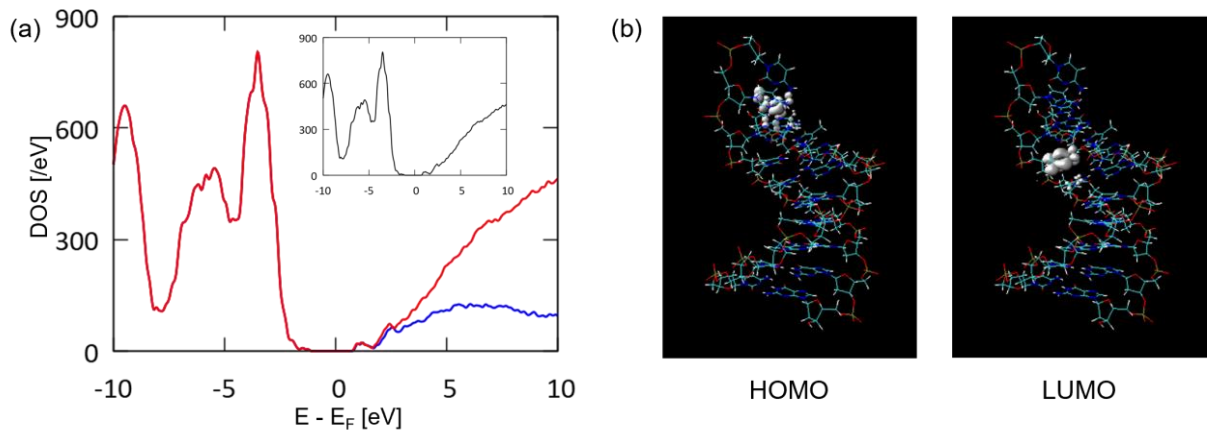

**Fig. 11**. (a) Density of states calculated using multi-site support functions ($r_{MS}$ = 4.2 Å) (blue, lower line), Sakurai-Sugiura method (red, upper line), and primitive PAOs (inset) and (b) molecular orbital pictures of hydrated DNA calculated using Sakurai-Sugiura method. (Reprinted with permission from Ref. 60. © 2017 American Chemical Society.)

*3.5 Metallic Nanoparticles*

In this section, we show an example of the calculations of metallic nanoparticles with MSSFs briefly. The size-controlled metallic nanoparticles show high catalytic reactivity, and the combination of nanoparticles and substrate is one of the important factors to affect the reactivity. The interface between the nanoparticle and the substrate is a kind of hyper-ordered structure. We have investigated an Au nanoparticle with 923 atoms in octahedral (*Oh*) symmetry, consisting of six layers (Fig. 12(a)), using DZP PAOs. The diameter of this six-layered nanoparticle is about 3 nm, which is close to the sizes used in actual experiments.[70)] The nano-size calculation model enables us to investigate the site-dependence of the atomic and electronic structures of nanoparticles. Figures 13(a) and 13(b) show the intra- and inter-layer nearest neighbor atomic distances of the nanoparticle, respectively. The atomic distances of the inner layers are close to those in a bulk fcc system, while they are distributed widely in the outer layers. The wide distribution of the intra-layer distances corresponds to the site dependence, i.e., the atomic distances around the center of the faces (about 2.9 Å) are longer than those around the vertices and edges (about 2.8 Å). Figures 14(a) and 14(b) show the projected DOS (pDOS) of an Au atom in a bulk system and at the vertex of the nanoparticle. The electronic structure at the vertex of the nanoparticle is quite different from that in the bulk, where the d-band center is shifted closer to the Fermi level, which suggests high reactivity at the vertex.

Not only the vertices of the nanoparticles but also the interface between the nanoparticles and the substrate have been considered as reaction active sites.[3] To treat nano-scale metallic nanoparticles (i.e., not clusters) on substrates, large calculation models with several thousand of atoms are required. With MSSFs, we can treat these large models of the catalytic systems. For example, Fig. 12(b) shows the optimized structures of Au nanoparticles on the MgO(001) substrate with 2,844 atoms in total, in which we removed the bottom part of the nanoparticle as found in experimental observations.[3,71] The detailed investigation of atomic and electronic structures and reactivity of metallic nanoparticles on the substrate will be provided in future studies.

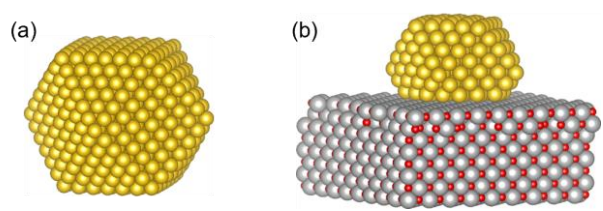

**Fig. 12**. Optimized structures of (a) Au nanoparticle in *Oh* symmetry with 923 atoms and (b) Au particle on Mg(001) surface.

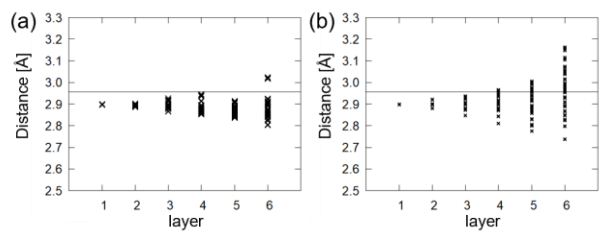

**Fig. 13**. (a) Intra-layer and (b) inter-layer nearest atom distances in Au nanoparticle with 923 atoms. Black horizontal lines correspond to the Au-Au distance in a bulk fcc gold (2.95 Å). Abscissa corresponds to the indices of the layers increasing from inner to outer of the nanoparticle. The sixth layer is the surface of the nanoparticle.

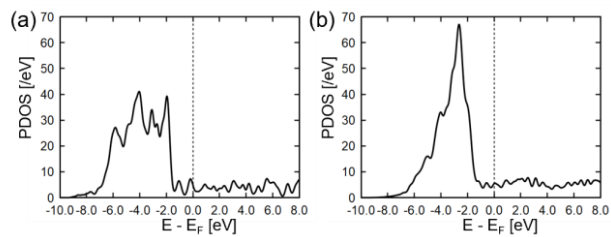

**Fig. 14**. Projected density of states (PDOS) of an Au atom (a) in bulk system and (b) at the vertex of an Au nanoparticle. Fermi levels are set to be zero (dashed line).

## 4. Conclusions

We have reviewed large-scale calculation methods, especially focusing on the methods in our large-scale density functional theory (DFT) code CONQUEST.[6–10] To model nano-scale complex structures, we often need large calculation models with several thousand atoms or more. DFT is a powerful tool to investigate the atomic and electronic structures of materials with high accuracy, but the computational cost of conventional DFT calculation methods is quite high (cubic scaling to system sizes). Therefore, special calculation techniques to treat such large systems are required.

The multi-site support function (MSSF) method[11,12] in CONQUEST makes it possible to improve both computational efficiency and accuracy. MSSFs are linear combinations of basis functions which belong not only to a target atom but also to its neighboring atoms, like local molecular orbitals (MOs). This MO-like picture of MSSFs enables us to reduce the number of the support functions to a minimal-basis size, while we can increase the number of the basis functions to improve computational accuracy, without increasing the number of the support functions. The linear-combination coefficients can be determined by using the local filter diagonalization method[55,56,11] and subsequent numerical optimization.[12] The investigations of accuracy for bulk Si and Al, hydrated DNA, bulk Fe and NiO demonstrate that MSSFs are applicable to varied materials such as insulating, semiconducting, metallic, and spin-polarized systems.

Examples of applications of MSSF to large systems with several thousand of atoms have been also shown. The geometry and electronic structure around complex interfaces were investigated using MSSF in these examples.[60–63] MSSF have been also applied to non-periodic materials such as biomolecules and metallic nanoparticle catalysts. The combination of MSSF and the Sakurai-Sugiura method,[68,69] an efficient interior eigenproblem solver, enable the MSSFs to be used to investigate the excited states of large systems.[60] Thus, we suggest that MSSF is now one of the most promising tools for investigation of hyper-ordered structures such as interfaces between nanoparticle catalysts and substrates.

**Acknowledgment**

This work is supported by the World Premier International Research Centre Initiative (WPI Initiative) on Materials Nanoarchitectonics (MANA), JSPS Grant-in-Aid for Transformative Research Areas (A) "Hyper-Ordered Structures Science" (Grant Nos. JP20H05883 and JP20H05878), JSPS Grant-in-Aid for Scientific Research (Grant No. JP18H01143), and JST

PRESTO (Grant No. JPMJPR20T4). This work is also partially supported by a project, JPNP16010, commissioned by the New Energy and Industrial Technology Development Organization (NEDO). Calculations were performed on the Numerical Materials Simulator at NIMS and the supercomputer HA8000 system at Kyushu University in Japan.

The authors are grateful for computational support from the UK Materials and Molecular Modeling Hub, which is partially funded by EPSRC (Grant No. EP/P020194), for which access was obtained via the UKCP consortium and funded by EPSRC (Grant Ref. No. EP/P022561/1). We also acknowledge computational support from the UK national high-performance computing service, ARCHER, for which access was obtained via the UKCP consortium and funded by EPSRC (Grant Ref. No. EP/P022561/1).

*E-mail: NAKATA.Ayako@nims.go.jp

**References**

1) S. Kohara, M. Shiga, Y. Onodera, H. Masai, A. Hirata, M. Murakami, T. Morishita, K. Kimura, and K. Hayashi, Sci. Rep. **11**, 22180 (2021).

2) Y. Takano, H. X. Kondo, Y. Kanematsu, and Y. Imada, Jpn. J. Appl. Phys. **59**, 010502 (2020).

3) H. Yoshida, Y. Kuwauchi, J. R. Jinschek, K. Sun, S. Tanaka, M. Kohyama, S. Shimada, M. Haruta, S. Takeda, Science **335**, 317 (2021).

4) M. Kawarasaki, K. Tanabe, I. Terasaki, Y. Fujii, and H. Taniguchi, Sci. Rep. **7**, 5351 (2017); K. Tsutsui and Y. Morikawa, Jpn. J. Appl. Phys. **59**, 010503 (2020).

5) R. M. Martin, *Electronic structure: basic theory and practical methods* (Cambridge University Press, 2004); R. O. Jones, Rev. Mod. Phys. **87**, 897 (2015).

6) Conquest website, http://www.order-n.org, (accessed February 2022).

7) E. Hernández, M. J. Gillan and C. M. Goringe, Phys. Rev. B: Condens. Matter Mater. Phys. **53**, 7147 (1996).

8) D. R. Bowler, R. Choudhury, M. J. Gillan, and T. Miyazaki, Phys. Status Solidi B **243**, 989 (2006).

9) D. R. Bowler and T. Miyazaki, J. Phys.: Condens. Matter **22**, 074207 (2010).

10) A. Nakata, J. S. Baker, S. Y. Mujahed, J. T. L. Poulton, S. Arapan, J. Lin, Z. Raza, S. Yadav, L. Truflandier, T. Miyazaki, and D. R. Bowler, J. Chem. Phys. **152**, 164112 (2020).

11) A. Nakata, D. R. Bowler, and T. Miyazaki, J. Chem. Theory Comput. **10**, 4813 (2014).

12) A. Nakata, D. R. Bowler, and T. Miyazaki, Phys. Chem. Chem. Phys. **17**, 31427 (2015).

13) M. Arita, S. Arapan, D. R. Bowler, and T. Miyazaki, J. Adv. Simulat. Sci. Eng. **1**, 87 (2014).

14) P. Hohenberg and W. Kohn, Phys. Rev. **136**, B864 (1964).

15) W. Kohn and L. J. Sham, Phys. Rev. **140**, A1133 (1965).

16) D. R. Bowler and T. Miyazaki, Rep. Prog. Phys. **75**, 036503 (2012).

17) W. Dawson, A. Degomme, M. Stella, T. Nakajima, L. E. Ratcliff, L. Genovese, WIREs. Comput. Mol. Sci. e1574 (2021); L. E. Ratcliff, S. Mohr, G. Huhs, T. Deutsch, M. Masella, L. Genovese, WIREs. Comput. Mol. Sci. **7**, e1290 (2017).

18) W. Yang, Phys. Rev. Lett. **66**, 1438 (1991).

19) W. Yang, Phys. Rev. A **44**, 7823 (1991).

20) W. Yang, J. Mol. Struct.: Theochem. **255**, 461 (1992).

21) G. Galli and M. Parrinello, Phys. Rev. Lett. **69** (1992) 3547 (1992).

22) E. Tsuchida, J. Phys. Soc. Jpn. **76**, 034708 (2007).

23) R. McWeeny, Rev. Mod. Phys. **32**, 335 (1960).

24) X. P. Li, R. W. Nunes and D. Vanderbilt, Phys. Rev. B: Condens. Matter Mater. Phys. **47**,

10891 (1993).

25) K. Kitaura, E. Ikeo, T. Asada, T. Nakano, and M. Uebayasi, Chem. Phys. Lett. **313**, 701 (1999).

26) J. M. Soler, E. Artacho, J. D. Gale, A. García, J. Junquera, P. Ordejón, and D. Sánchez-Portal, J. Phys. Cond. Matt. **14**, 2745 (2002).

27) A. García, N. Papior, A. Akhtar, E. Artacho, V. Blum, E. Bosoni, P. Brandimarte, M. Brandbyge, J. I. Cerdá, F. Corsetti, R. Cuadrado, V. Dikan, J. Ferrer, J. Gale, P. García-Fernández, V. M. García-Suárez, S. García, G. Huhs, S. Illera, R. Korytár, P. Koval, I. Lebedeva, L. Lin, P. López-Tarifa, S. G. Mayo, S. Mohr, P. Ordejón, A. Postnikov, Y. Pouillon, M. Pruneda, R. Robles, D. Sánchez-Portal, J. M. Soler, R. Ullah, V. W-z Yu, and J. Junquera, J. Chem. Phys. **152**, 204108 (2020).

28) T. Ozaki, Phys. Rev. B. **67**, 155108, (2003).

29) S. Tsuneyuki, T. Kobori, K. Akagi, K. Sodeyama, K. Terakura, and H. Fukuyama, Chem. Phys. Lett. **476**, 104 (2009).

30) G. M. J. Barca, C. Bertoni, L. Carrington, D. Datta, N. De Silva, J. E. Deustua, D. G. Fedorov, J. R. Gour, A. O. Gunina, E. Guidez, T. Harville, S. Irle, J. Ivanic, K. Kowalski, S. S. Leang, H. Li, W. Li, J. J. Lutz, I. Magoulas, J. Mato, V. Mironov, H. Nakata, B. Q. Pham, P. Piecuch, D. Poole, S. R. Pruitt, A. P. Rendell, L. B. Roskop, K. Ruedenberg, T. Sattasathuchana, M. W. Schmidt, J. Shen, L. Slipchenko, M. Sosonkina, V. Sundriyal, A. Tiwari, J. L. G. Vallejo, B. Westheimer, M. Wloch, P. Xu, F. Zahariev, and M. S. Gordon, J. Chem. Phys. **152**, 154102 (2020).

31) S. Tanaka, Y. Mochizuki, Y. Komeiji, Y. Okiyama, and K. Fukuzawa, Phys. Chem. Chem. Phys. **16**, 10310 (2014).

32) E. Tsuchida and M. Tsukada, J. Phys. Soc. Jpn. 67, 3844 (1998).

33) J. C. A. Prentice, J. Aarons, J. C. Womack, A. E. A. Allen, L. Andrinopoulos, L. Anton, R. A. Bell, A. Bhandari, G. A. Bramley, R. J. Charlton, R. J. Clements, D. J. Cole, G. Constantinescu, F. Corsetti, S. M.-M. Dubois, K. K. B. Duff, J. M. Escartín, A. Greco, Q. Hill, L. P. Lee, E. Linscott, D. D. O'Regan, M. J. S. Phipps, L. E. Ratcliff, Á. Ruiz Serrano, E. W. Tait, G. Teobaldi, V. Vitale, N. Yeung, T. J. Zuehlsdorff, J. Dziedzic, P. D. Haynes, N. D. M. Hine, A. A. Mostofi, M. C. Payne, and C.-K. Skylaris, J. Chem. Phys. **152**, 174111 (2020).

34) M. Ceriotti, T. D. Kühne, and M. Parrinello, J. Chem. Phys. **129**, 024707 (2008).

35) L. E. Ratcliff, W. Dawson, G. Fisicaro, D. Caliste, S. Mohr, A. Degomme, B. Videau, V. Cristiglio, M. Stella, M. D'Alessandro, S. Goedecker, T. Nakajima, T. Deutsch, and L.

Genovese, J. Chem. Phys. **152**, 194110 (2020).

36) A. M. N. Niklasson, Phys. Rev. B **66** 155115 (2002).

37) E. Rudberg, E. H.Rubensson, P. Sałek, A. Kruchinina, SoftwareX **7**, 107, (2018).

38) T. D. Kühne, M. Iannuzzi, M. D. Ben, V. V. Rybkin, P. Seewald, F. Stein, T. Laino, R. Z. Khaliullin, O. Schütt, F. Schiffmann, D. Golze, J. Wilhelm, S. Chulkov, M. H. Bani-Hashemian, V. Weber, U. Borštnik, M. Taillefumier, A. S. Jakobovits, A. Lazzaro, H. Pabst, T. Müller, R. Schade, M. Guidon, S. Andermatt, N. Holmberg, G. K. Schenter, A. Hehn, A. Bussy, F. Belleflamme, G. Tabacchi, A. Glöß, M. Lass, I. Bethune, C. J. Mundy, C. Plessl, M. Watkins, J. VandeVondele, M. Krack, and J. Hutter, J. Chem. Phys. **152**, 194103 (2020).

39) J. Junquera, Ó. Paz, D. Sánchez-Portal, and E. Artacho, Phys. Rev. B **64**, 235111 (2001).

40) E. Anglada, J. M. Soler, J. Junquera, and E. Artacho, Phys. Rev. B **66**, 205101 (2002).

41) T. Ozaki, Phys. Rev. B **67**, 155108 (2003).

42) T. Ozaki and H. Kino, Phys. Rev. B **69**, 195113 (2004).

43) A. S. Torralba, M. Todorović, V. Brázdová, R. Choudhury, T. Miyazaki, M. J. Gillan, D. R. Bowler, J. Phys.: Condens. Matter **20**, 294206 (2008).

44) D. R. Bowler, J. S. Baker, J. T. L. Poulton, S. Y. Mujahed, J. Lin, S. Yadav, Z. Raza and T. Miyazaki, Jpn. J. Appl. Phys. 58, 100503 (2019).

45) E. Hernández, M. J. Gillan, and C. M. Goringe, Phys. Rev. B: Condens. Matter Mater. Phys., **55**, 13485 (1997).

46) A. A. Mostofi, P. D. Haynes, C.-K. Skylaris, and M. C. Payne, J. Chem. Phys. **119**, 8842 (2003).

47) S. Goedecker, Wavelets and Their Application: For The Solution of Partial Differential Equations in Physics (Presses Polytechniques et Universitaires Romandes, 1998).

48) J. VandeVondele, M. Krack, F. Mohamed, M. Parrinello, T. Chassaing, and J. Hutter, Comput. Phys. Commun. **167**, 103 (2005).

49) J. Iwata, D. Takahashi, A. Oshiyama, T. Boku, K. Shiraishi, S. Okada, K. Yabana, J. Comput. Phys. **229**, 2339 (2010).

50) J. R. Chelikowsky, N. Troullier, and Y. Saad, Phys. Rev. Lett. **72**, 1240 (1994).

51) D. R. Hamann, Phys. Rev. B **88**, 085117 (2013).

52) D. Vanderbilt, Phys. Rev. B **32**, 8412 (1985).

53) ATOM code for the generation of norm-conserving pseudopotentials. The version maintained by the SIESTA project, http://icmab.es/siesta/Pseudopotentials/index.html, (accessed February 2022).

54) E. Hernández and M. J. Gillan, Phys. Rev. B **51**, 10157 (1995).

55) M. J. Rayson and P. R. Briddon, Phys. Rev. B **80**, 205104 (2009).

56) M. J. Rayson, Comput. Phys. Commun. **181**, 1051 (2010).

57) T. Miyazaki, D. R. Bowler, R. Choudhury, and M. J. Gillan, J. Chem. Phys. **121**, 6186 (2004).

58) J. P. Perdew, A. Zunger, Phys. Rev. B **23**, 5048 (1981).

59) J. P. Perdew, K. Burke, M. Ernzerhof, Phys. Rev. Lett. **77**, 3865 (1996).

60) A. Nakata, Y. Futamura, T. Sakurai, D. R. Bowler, and T. Miyazaki, J. Chem. Theory Comput. **13**, 4146 (2017).

61) C. Romero-Muñiz, A. Nakata, P. Pou, D. R. Bowler, T. Miyazaki, and R. Pérez, J. Phys.: Condens. Matter **30**, 505901 (2018).

62) D. R. Småbråten, A. Nakata, D. Meier, T. Miyazaki, and S. M. Selbach, Phys. Rev. B **102**, 144103 (2020).

63) J. S. Baker and D. R. Bowler, Adv. Theory Simul. **3**, 2000154 (2020).

64) P. E. Blöchl, Phys. Rev. B **50**, 17953 (1994).

65) G. Kresse and J. Furthmüller, Phys. Rev. B **54**, 11169 (1996).

66) G. Kresse and D. Joubert, Phys. Rev. B **59**, 1758 (1999).

67) S. Grimme, J. Comput. Chem. **27**, 1787 (2006).

68) T. Sakurai and H. Sugiura, J. Comput. Appl. Math. **159**, 119 (2003).

69) T. Sakurai, Y. Futamura, and H. Tadano, J. Algorithms Comput. Technol **7**, 249 (2013).

70) T. Mitsudome, A. Noujima, T. Mizugaki, K. Jitsukawa, K. Kaneda, Chem. Commun. **2009**, 5302 (2009); T. Urayama, T. Mitsudome, Z. Maeno, T. Mizugaki, K. Jitsukawa, and K. Kaneda, Chem. Lett. **44**, 1062 (2015).

71) S. Takeda, Y. Kuwauchi, H. Yoshida, Ultramicroscopy **151**, 178 (2015).